\documentclass[aps, pra, twocolumn, english, showpacs, amsmath, amssymb]{revtex4-1}

\usepackage{graphicx}
\usepackage{dcolumn}
\usepackage{bm}
\usepackage[utf8]{inputenc}
\usepackage{default}
\usepackage{braket}
\usepackage{amsmath}
\usepackage{xcolor}
\usepackage[version=3]{mhchem}
\usepackage{bbold}
\usepackage{mathtools}
\usepackage{float}
\usepackage{babel}

\newcommand{\WW}{\mathcal{W}}

\DeclareMathOperator{\Tr}{Tr}

\graphicspath{{pix/}}

\begin{document}


\title{Proving genuine multiparticle entanglement \\
from separable nearest-neighbor marginals}

\author{Marius Paraschiv}
\author{Nikolai Miklin}
\author{Tobias Moroder}
\author{Otfried G{\"u}hne}
\affiliation{%
Naturwissenschaftlich-Technische Fakult{\"a}t, 
Universit{\"a}t Siegen, Walter-Flex-Stra{\ss}e 3, 57068 Siegen, Germany
}%
\date{\today}

\begin{abstract}
We address the question of whether or not global entanglement of a 
quantum state can be inferred from local properties. Specifically, 
we are interested in genuinely multiparticle entangled states whose 
two-body marginals are all separable, but where the entanglement can be 
proven using knowledge of a subset of the marginals only. Using an 
iteration of semidefinite programs we prove that for any possible 
marginal configuration up to six particles multiqubit states with 
the desired properties can be found. We then present a method to 
construct states with the same properties for more particles in 
higher dimensions.
\end{abstract}

\maketitle


\section{Introduction}
\label{sec:level1}

An essential property of quantum systems is that they can be entangled, 
meaning that the state of the system cannot be factorized 
\cite{hreview, gtreview}. A related question concerns the relationship 
between global properties of the system and the local properties of its 
subsystems. In the simplest version, one may just ask whether the global 
state can be determined from its marginals, and which sets of marginals 
are compatible. This is an rather old problem, sometimes called the marginal 
problem, or the representability problem \cite{coleman}, but recently it
attracted again much attention \cite{lpw2002, diosi2004, joneslinden2005, 
walter2013, sawicki2013, schilling2017, klyachko2004, wyderka2017}. More
precisely, one can also ask whether certain global properties, such as 
entanglement, can be concluded from the marginals. In fact, several 
examples have been identified, where this is the case: Using spin-squeezing
inequalities one may prove entanglement from two-body marginals, although
these marginals itself are separable \cite{toth1, toth2}. Similar phenomena
exist for Bell inequalities, where the marginals are compatible with a local 
hidden-variable model, but the global state is not, and this can be proven 
from the marginals \cite{wurflinger, tura, wang}.

The notion of entanglement used in the above mentioned works relies on the
question whether or not the global state can be factorized completely, 
that is, whether it is fully separable or not. In other words, if the 
state does not factorize, it is entangled. This does not mean, however, 
that it is genuine multiparticle entangled, as genuine multiparticle entanglement
requires the entanglement between all particles and not only some of them. So the 
more demanding task of proving that a state is genuinely multiparticle entangled 
just from separable two-body marginals still remained. In Ref.~\cite{marco}, however, 
a first example of this phenomenon has been presented and in Ref.~\cite{nikolai} 
the authors provided a systematic method of finding genuinely multiparticle entangled 
states with separable two-body marginals, where the genuine multiparticle entanglement 
can be proven from the marginals only. They have also provided a scheme for constructing 
states with the desired properties for any number of particles, and gave examples 
of numerically found states for up to five particles.

\begin{figure}[t]
  \begin{center}
    \includegraphics[width=0.7\columnwidth]{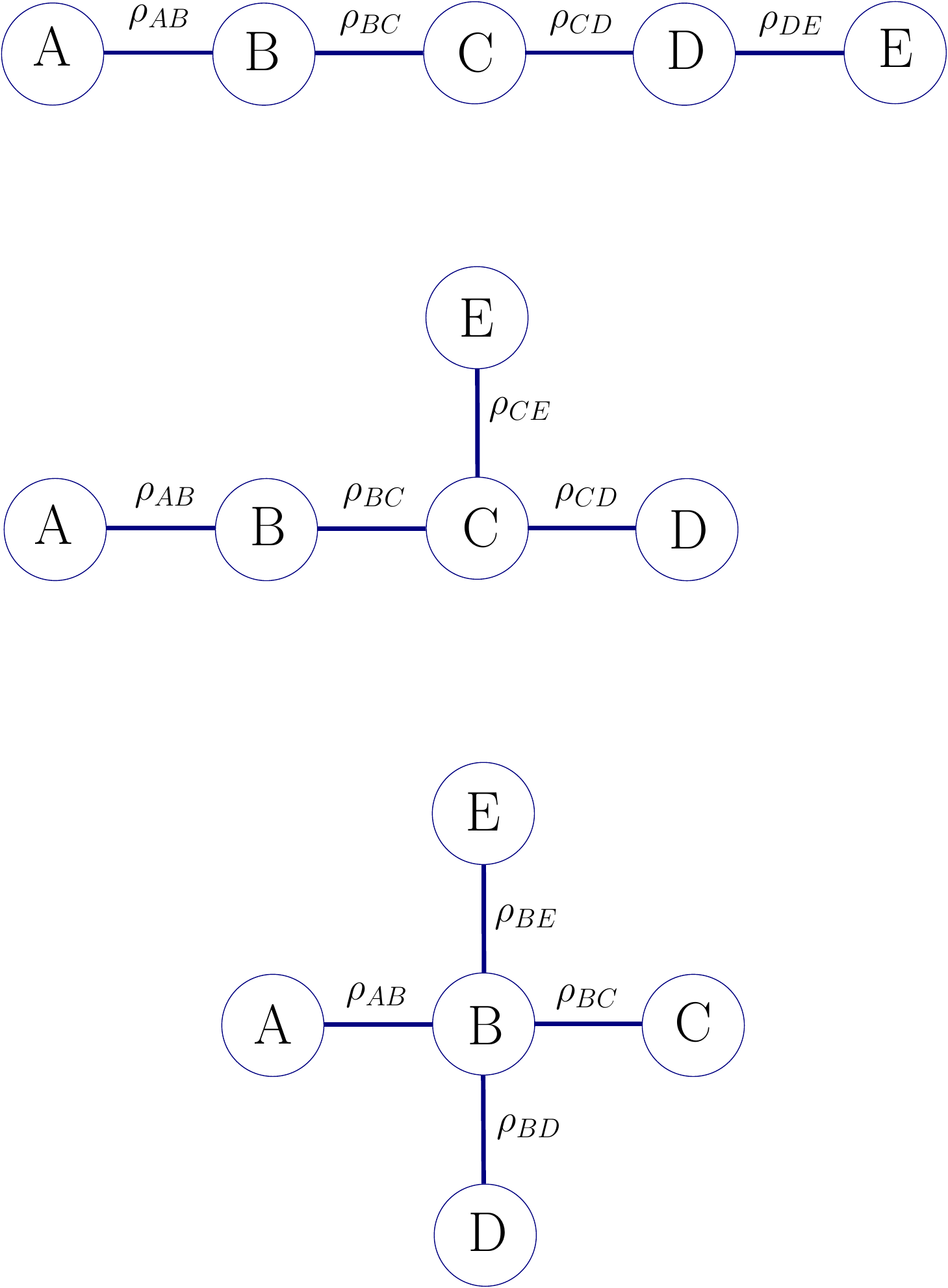}
    \end{center}
    \caption{If global entanglement should be concluded from a set of two-body marginals $\{\rho_{ij}\}$
    then the set of marginals has to obey two conditions: Firstly, any particle $k$ must be covered by the set 
    and secondly, the graph arising from the interpretation of the $\rho_{ij}$ as edges must be connected. 
    Among the marginal sets with these conditions, the minimal sets are of special interest, and these are
    necessarily tree-like configurations. The figure shows all minimal tree configurations of marginals 
    for five particles, up to permutations of the particles \cite{steinbach}.}
  \label{fig:5qbconf}
\end{figure}


In this paper we  go one step further and see what happens if only 
a subset of the marginals is known, but still all marginals are required 
to be separable. Using a suitable ordering and topology of the particles, 
one can always view the subset of known marginals as the set of nearest-neighbor 
marginals.  It is clear that only from subsets of two-body 
marginals where all particles are included and where all marginals form
a connected graph the global entanglement can be proven. The possible 
marginal configurations are known as unlabeled trees \cite{steinbach}. 
The number of these configurations is known to scale exponentially \cite{otter}, 
but for a small number of particles an exhaustive classification is 
known \cite{steinbach, oeis}. We will see that it is always possible to 
find examples of states where genuine multiparticle entanglement can be proven from separable two-body 
nearest-neighbor marginals.

In order to study different marginal configurations for higher particle 
numbers the numerical tools used in Ref.~\cite{nikolai} are not sufficient 
and improved optimization methods are required. The extension to more 
particles, however, gives new insights: for the minimal configurations 
of five (see Fig.~\ref{fig:5qbconf}) and six qubits, examples of states 
can always be found already for qubits and, remarkably, most of these 
states are pure. This purity allows to present a method to find examples 
for general marginal configurations of an arbitrary number of particles 
using copies of the numerically found five- and six-qubit states. This 
method allows one to construct the desired states for any marginal 
configuration of an arbitrary number of particles, but 
higher-dimensional systems are required.

This paper is organized as follows. In Section II we introduce the required facts about multiparticle
entanglement and state the problem. In Section III we describe the iteration of semidefinite programs
that we use to solve this problem. Section IV presents the results for four, five and six qubits. In
Section V we discuss the generalization to an arbitrary number of particles. Finally, we conclude 
and discuss further research directions.


\section{Definitions and statement of the problem}
\label{sec:level1} 

We begin by recalling the notion of genuinely multiparticle entangled states, 
detailed discussions can be found in Refs.~\cite{hreview, gtreview, bastian}. 
For simplicity, we will restrict ourselves to three-particle systems, but the definitions 
are valid for an arbitrary number of particles. First, a state $\rho_{ABC}$ is said 
to be {separable with respect to a bipartition} $A|BC$ if it can be written 
as a mixture of product states, with respect to the bipartition $A|BC$
\begin{equation}
\label{eq1}
\rho^{\rm{sep}}_{A|BC} = 
\sum_{k} q_k \Ket{\phi^k_A}\Bra{\phi^k_A} \otimes \Ket {\psi^k_{BC}} \Bra {\psi^k_{BC}},
\end{equation}
where the $q_k$ form a probability distribution. If the global state of the system can be 
written as
\begin{equation}
\label{eq2}
\rho^{\rm{bs}} = 
p_{1} \rho^{\rm{sep}}_{A|BC} + p_{2} \rho^{\rm{sep}}_{B|AC} + p_{3} \rho^{\rm{sep}}_{C|AB}
\end{equation}
it is called {biseparable}. This gives the definition of genuine multiparticle entanglement: 
If a state is not biseparable, i.e. it cannot be written in the form of Eq.~(\ref{eq2}), 
then it is genuinely multiparticle entangled.

\begin{figure}
  \centering
    \includegraphics[width=0.9\columnwidth]{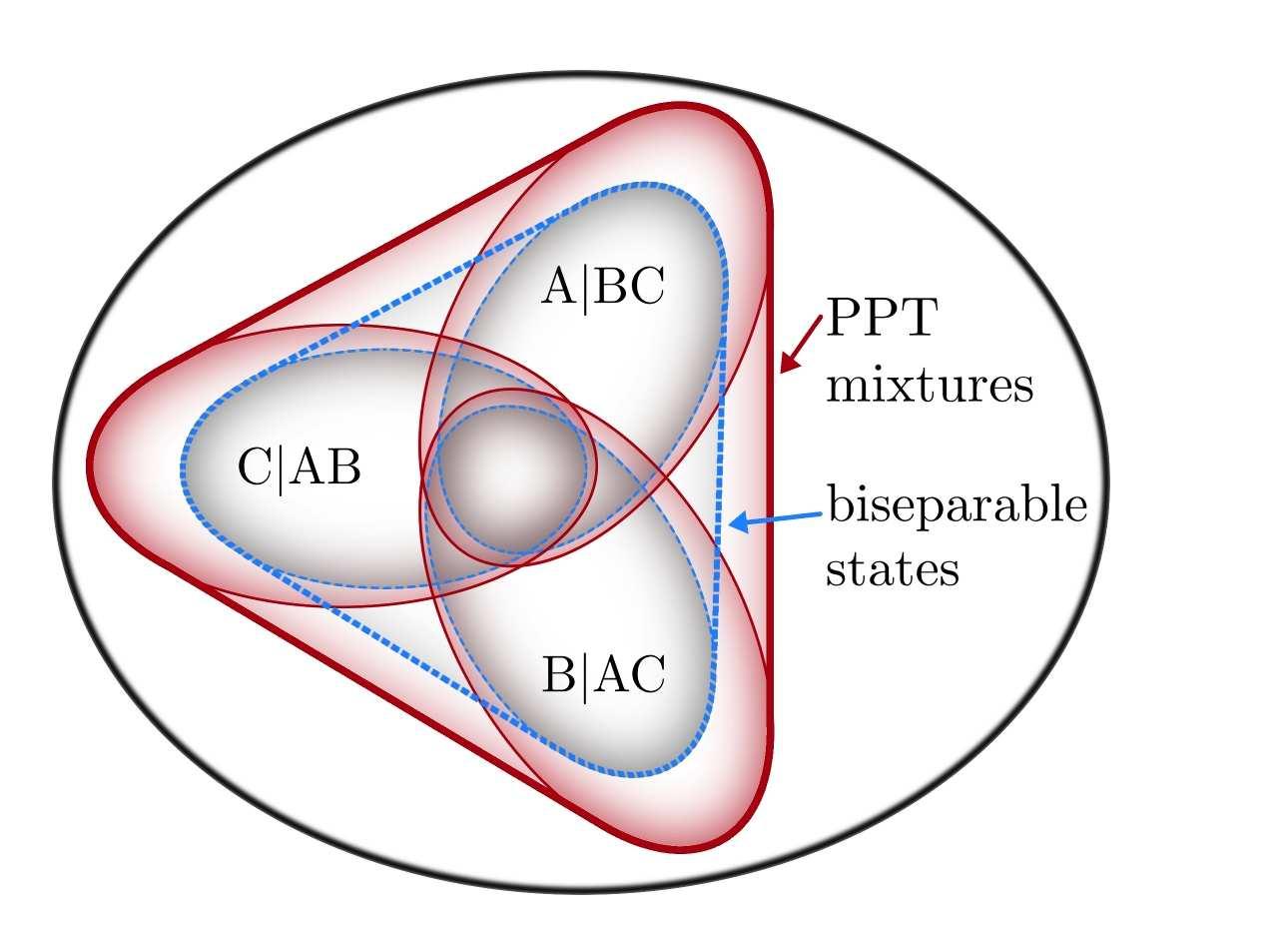}
    \caption{Illustration of biseparable states and PPT mixtures for a three-particle system, see text
    for further details. The figure is taken from Ref.~\cite{bastian}.}
  \label{fig:pptmix}
\end{figure}

Due to the definition of biseparability, it is very difficult to verify this property 
directly. For our approach it is very useful to consider a relaxed definition by considering 
a larger set of states, the set of so-called PPT mixtures (see Fig.~\ref{fig:pptmix}). Let us first
recall the entanglement criterion of the positivity of the partial transpose (PPT). Any 
two-particle state on an $N\times M$-system can be written as
\begin{equation}
\label{eq3}
\rho = \sum_{i,j}^{N}\sum_{k,l}^{M} \rho_{ij,kl}\Ket{i}\Bra{j}\otimes\Ket{k}\Bra{l}.
\end{equation}
The partial transposition of $\rho$ with respect to the first subsystem (we use the standard convention 
of naming the two subsystems Alice and Bob), is then given by
\begin{equation}\label{eq4}
\rho^{T_A}=\sum_{i,j}^{N}\sum_{k,l}^{M}\rho_{ji ,kl}\Ket{i}\Bra{j}\otimes\Ket{k}\Bra{l}.
\end{equation}
A state $\rho$ is said to have a positive partial transpose (PPT) if
\begin{equation}
\label{eq5}
\rho^{T_A} \geq 0,
\end{equation}
that is, $\rho^{T_A}$ has no negative eigenvalues. Separable states are PPT \cite{peres} 
and according to the Horodecki theorem \cite{horodeki}, for $2 \times 2$ and $2 \times 3$ 
systems, any PPT state is also separable. This criterion for separability is very easy to 
test numerically, thus we shall use it to test the separability of our two-body marginals. 

For multiparticle states, the partial transposition can be defined for any bipartition of
the system. Now, similarly to biseparable states, a state that can be written as
\begin{equation}\label{eq6}
\rho^{\rm{pmix}} = p_{1} \rho^{\rm{ppt}}_{A|BC} + p_{2} \rho^{\rm{ppt}}_{B|AC} + p_{3} \rho^{\rm{ppt}}_{C|AB}
\end{equation}
is called a PPT mixture, as it is a mixture of PPT states for the different bipartitions 
\cite{bastian}.

Looking at Fig.~\ref{fig:pptmix}, the convex hull of all states separable with respect to a fixed 
bipartition is the set of biseparable states. In a similar way, the convex hull of states which are 
PPT with respect to a bipartition is the set of PPT mixtures. It is clear that every biseparable state 
is also a PPT mixture. Thus, if we can prove that a state is not a PPT mixture, then it is 
genuinely multiparticle entangled. Note that the partial transposition is only one example 
of a map that can be used, other positive but not completely positive maps work as well
\cite{cecilia}.


Now, having a suitable criterion for entanglement, we can write down an entanglement
witness that can 
detect a state which is not a PPT mixture. An entanglement witness is an observable $\WW$
that is non-negative on all biseparable states and has a negative expectation value on at 
least one entangled state. For the two-particle case a witness $\WW$ is called {decomposable} 
if it can be written in terms of two positive semidefinite observables $P$ and $Q$ ($P\geq 0$ 
and $Q\geq 0$) as
\begin{equation}\label{eq7}
\WW = P + Q^{T_A}.
\end{equation}
One can generalize this definition to the multiparticle case. A witness that can be written as
\begin{equation}\label{eq8}
\WW = P_M + Q_M^{T_M},
\end{equation}
for {\it any} bipartition $M|\bar{M}$ of the system is called {fully decomposable}. The connection
to the notion of PPT mixtures is the following:

\textbf{Observation 1.} 
{\it If $\rho$ is not a PPT mixture, then there exists a fully decomposable witness $\WW$ 
that detects it. The proof can be found in Ref.~\cite{bastian}.}

Now we can define the problem in a rigorous way. For $N$ particles there are $N(N-1)/2$ 
possible two-body marginals (reduced density matrices) $\rho_{ij}$. We fix a subset $S$ of 
them and call them also the nearest-neighbor marginals. Then, we want to find an $N$-particle state $\rho$ such that:
\begin{enumerate}
\item All $N(N-1)/2$ two-body marginals of $\rho$ are separable. Since we are 
first looking for multi-qubit states, the marginals are systems of two qubits 
and for them separability is equivalent to being PPT. 
\item 
The state $\rho$ is genuinely multiparticle entangled and this entanglement can be 
proven from knowledge of the marginals in the subset $S$ only. This condition can 
be assured by using fully decomposable witnesses, which detect states that are 
not PPT-mixtures and which contain only two-body interactions corresponding to 
the marginals in the subset~$S$.
\end{enumerate}
Clearly, the subset $S$ has to obey some conditions, in order to find the requested states.
Firstly, any particle $k$ must be covered by the set and secondly, the graph arising from 
the interpretation of the $\rho_{ij}$ as edges must be connected, otherwise there is one
bipartition, for which entanglement cannot be checked (see also Fig.~\ref{fig:5qbconf}). 
Among the marginal sets with these conditions, the minimal sets are of special interest as in these
the least amount of knowledge is given. These configurations are
 necessarily tree-like configurations. 

\section{Description of the algorithm}
\label{sec:level2}
In this section, we describe the algorithm used for finding the 
desired states. The algorithm relies on an iteration of semidefinite 
programs (SDPs) \cite{boyd}, some basic facts about SDPs are explained 
in the Appendix A.


To obtain a state with the desired properties, we implement a 
program as a sequence of steps, over which an iteration is 
performed until the desired precision is reached. The main
idea of this program was already used in Ref.~\cite{nikolai}.

\textbf{Step 0}: Generate a random pure state $\rho$. For practical
purposes, it is preferable that the initial state does not have any 
symmetries. Otherwise, the following iteration may in practice end
up in a fixed point which does not have the desired properties.

\textbf{Step 1}: Insert the state into the first SDP. 
This SDP aims at finding an optimal fully decomposable witness, meaning 
that the witness has the smallest expectation value (among all other 
considered witnesses) for the given state. The witness is constructed such 
that it can be evaluated from knowledge of the marginals $\rho_{ij}$ in
the subset $S$ only. Formally, this program is given by:
\begin{equation}
\label{eq9}
\begin{split}
    \min & \operatorname{Tr}(\WW \rho) \\
    \rm{s.t.} & \operatorname{Tr}(\WW) = 1, \\
    & \WW = \sum_{i,j} \omega_{i,j}^{\alpha, \beta} \sigma_i^{\alpha} \otimes \sigma_j^{\beta} \otimes \mathbb{1}^{\otimes(N-2)} + {\rm perm.\; in}\; S, \\
    & \WW = P_M + Q_M^{T_M} \;{\rm with}\;  P_M,Q_M \geq 0 \; {\rm for\; all} \; M \vert \bar{M}.
    \\
\end{split}
\end{equation}

The first condition, $Tr(\WW)=1$ is just a normalization condition on the 
witness. This normalization is not the only possible one, it does, however, 
assure the best robustness against white noise \cite{bastian}. 

The second condition ensures the constraint that the witness contains only contains two-body 
terms from the marginals in $S$. Here, $\sigma_i^\alpha$ denotes a Pauli matrix acting on
the qubit $\alpha$. The permutation is performed over two-body marginals within
this set, with a different $\omega_{i,j}^{\alpha, \beta}$ coefficient for each term. It is 
also important to stress here that, while the witness is restricted to only two-body terms, 
the operators $P_M$ and $Q_M$ are not. 

Finally, the last condition ensures that the witness is fully decomposable, hence it 
detects non-PPT mixtures.

\textbf{Step 2}: 
Insert the optimal witness from the previous step, into the second SDP.
The purpose of this SDP is to obtain an optimal state that minimizes 
the expectation value of $\WW$ as much as possible under the condition 
that all two-body marginals are PPT. This program is:
\begin{equation}\label{eq10}
\begin{split}
  \min  & \operatorname{Tr}(\WW \rho) \\
  \rm{s.t.} & \operatorname{Tr}(\rho) = 1. \\
  & \rho \geq 0, \\
  & \rho_{\alpha, \beta}^{T_\alpha} \geq 0 \;{\rm for \; all}\; \alpha, \beta.\\
\end{split}
\end{equation}
While the first two conditions are present just to assure the fact that $\rho$ is a density matrix, 
the third one represents the constraint that all two-body marginals of $\rho$ must be separable 
(for qubits the PPT condition is equivalent to separability). Separability must hold for all 
two-body marginals, not just the marginals in $S$.

One can then iterate the steps 1 and 2, obtaining a better approximation of the 
optimal state with each additional step. This see-saw algorithm is, of course,
not guaranteed to converge to the global optimum. However, in practice different
solutions turned out to be equivalent under local unitary transformations. 

In practice, the two SDPs have been implemented in Python, using the Picos 
convex optimization interface \cite{picos}. After a remarkably small number of iterations (usually two 
or three) one already finds a state that satisfies the desired requirements. On a regular 
desktop configuration, the four-qubit state is obtained in less than a minute, the five-qubit 
state in around 45 minutes and the six-qubit state in around 6 hours. We thus managed to obtain 
such states for four, five and six qubits, for various configurations (see Table I below). We 
started with a pure initial random state and it is important to mention that all obtained optimal 
states are also pure, except for two of them, see Table I. The obtained pure states are also uniquely determined by the marginals $\rho_{i,j}$ in $S$, which can be concluded from the fact that they are 
eigenstates corresponding to the smallest eigenvalue of the witness, and this eigenvalue
is non-degenerate. This will turn out to be essential later on, when we discuss generalizations 
of these results.

The negative expectation values found numerically (and presented in Table I) may seem small 
($10^{-4}$ for six-qubit states) but it is important to note that the SDP solver works with 
a precision of at least $10^{-10}$. 

One can observe that, as the number $N$ of particles increases, the absolute value 
of the witness value for the numerically-found state decreases. This may be expected, 
since the total number of two-body marginals is ${{N}\choose{2}}=N(N-1)/2$ and the total
number of parameters of the density matrix is $4^N-1$. On the other hand, the number of 
nearest-neighbor two-body marginals is $N-1$ and in total $3(N-1)$ parameters of the
density matrix are known. Thus, as $N$ grows, the  witness has information about a very small fraction of all marginals and parameters and the phenomenon becomes fragile.

\section{Results for the various configurations}
\label{sec:level3}

In this section, we present the results obtained by using the previously 
described method. After finding the numerical form of the desired states, 
one also needs to find an analytical approximation. This is a tedious task 
and has only been done here for the four- and five-qubit states. By noting 
that local unitary transformations do not affect the entanglement in our system, 
we can apply local unitary transformations to the state
\begin{equation}
\rho_{\rm{num.}}(\alpha, \theta, \phi) = U^{\dagger} \rho_{\rm{num.}} U
\end{equation}
and perform an optimization over any function of the parameters $\alpha, \theta$ 
and $\phi$. This generally leads to a state $\rho$ with some zero elements, simplifying 
the task of finding an analytical form. The above local unitaries are constructed from 
tensor products of qubit unitaries, which can be parametrized as
\begin{equation}
U(\alpha,\theta, \phi) = \left(
\begin{array}{cc}
 e^{i \alpha } \cos (\phi ) & e^{i \theta } \sin (\phi ) \\
 -e^{-i \theta } \sin (\phi ) & e^{-i \alpha } \cos (\phi ) \\
\end{array}
\right)
\end{equation}
The main results of the following discussion are summarized in Table~I.

\begin{table}[t]
\begin{tabular}{|c|c|c|c|}
\hline
\multicolumn{1}{|c|}{\textbf{ Index \hspace{0.1cm} }} & \multicolumn{1}{c|}{\textbf{ Configuration \hspace{0.1cm} }} & \multicolumn{1}{c|}{\textbf{\hspace{0.2cm} $ \Tr[\rho \WW] $ \hspace{0.2cm} }} & \multicolumn{1}{c|}{\textbf{ \hspace{0.1cm}  pure \hspace{0.1cm} }}\tabularnewline
\hline
\hline
\multicolumn{4}{|c|}{\textbf{4-qubit configurations}}\tabularnewline
\hline
\hline
 4a & \includegraphics[scale=0.35]{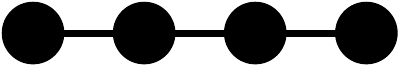} & $-3.15 \cdot 10^{-3}$ & No \tabularnewline
\hline
 4b & \includegraphics[scale=0.35]{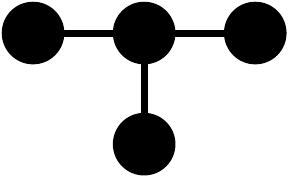} & $-3.56\cdot 10^{-3}$ & Yes\tabularnewline
\hline
\hline
\multicolumn{4}{|c|}{\textbf{5-qubit configurations}}\tabularnewline
\hline
\hline
 5a &  \includegraphics[scale=0.35]{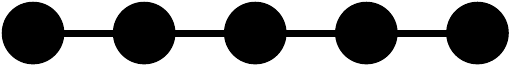} & $-1.13 \cdot 10^{-3}$ & Yes\tabularnewline
\hline
 5b &  \includegraphics[scale=0.35]{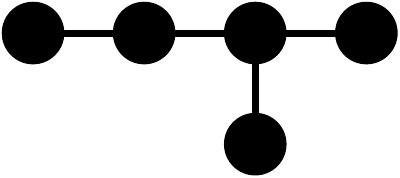} & $-1.31 \cdot 10^{-3}$ & Yes\tabularnewline
\hline
 5c &  \includegraphics[scale=0.35]{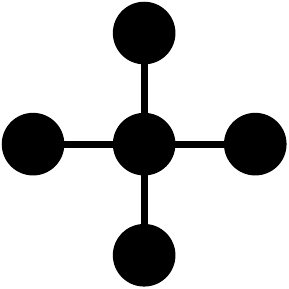} & $-1.38 \cdot 10^{-3}$ & Yes\tabularnewline
\hline
\hline
\multicolumn{4}{|c|}{\textbf{6-qubit configurations}}\tabularnewline
\hline
\hline
 6a &  \includegraphics[scale=0.35]{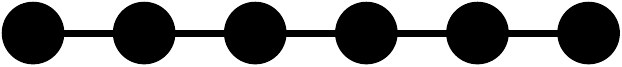} & $-2.01 \cdot 10^{-4}$ & Yes\tabularnewline
\hline
 6b &  \includegraphics[scale=0.35]{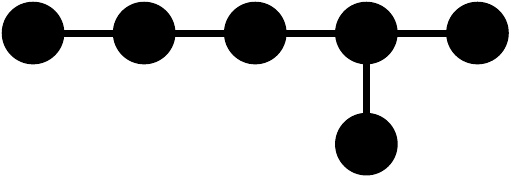} & $-2.56 \cdot 10^{-4}$ & Yes\tabularnewline
\hline
 6c &  \includegraphics[scale=0.35]{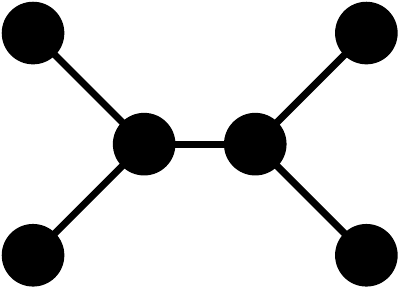} & $-2.84 \cdot 10^{-4}$ & Yes\tabularnewline
\hline
 6d &  \includegraphics[scale=0.35]{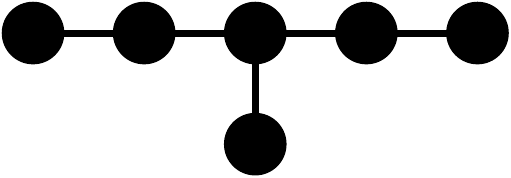} & $-2.92 \cdot 10^{-4}$ & Yes\tabularnewline
\hline
 6e &  \includegraphics[scale=0.35]{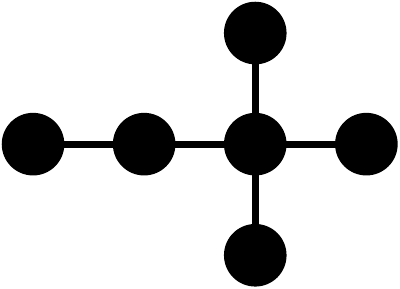} & $-3.80 \cdot 10^{-4}$ & Yes\tabularnewline
\hline
 6f &  \includegraphics[scale=0.35]{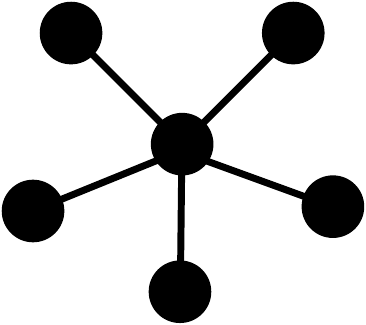} & $-4.54 \cdot 10^{-4}$ & No\tabularnewline
\hline
\end{tabular}
\caption{Obtained states and witness values for the various configurations. All states are 
uniquely determined by their known two-body marginals from the set $S$.}
\end{table}

\subsection{Four qubits}
\label{sec:level21}

There are two possible configurations for four qubits, denoted in Table I 
as 4a and 4b. While for both configurations we obtained genuinely multiparticle 
entangled states with the desired requirements, state 4a is not pure. We 
show here the other state, 4b, which is pure and uniquely determined by 
its known two-body marginals.
\begin{equation}
\label{eq11}
\Ket{\psi_4} = \frac{1}{\sqrt{87}}(5\Ket{\phi_1} +\sqrt{10}\Ket{\phi_2} + \sqrt{3}\Ket{\phi_3} + 7\Ket{\phi_4})
\end{equation}
where the component states are
\begin{align}
\Ket{\phi_1} = &\frac{1}{\sqrt{2}}(e^{\frac{3\pi}{7}i} \Ket{1100} - \Ket{0000})
\nonumber
\\
\Ket{\phi_2} = &\frac{1}{\sqrt{5}}(e^{\frac{\pi}{4}i} \Ket{0101} - \Ket{0111} - \Ket{1000}-
\nonumber
\\
&- \Ket{1001} - \Ket{1111})
\nonumber
\\
\Ket{\phi_3} = &\frac{1}{\sqrt{6}}(e^{\frac{-2\pi}{3}i} \Ket{0010} + \Ket{0011} + \Ket{0100} + \Ket{1010} +
\nonumber
\\
&+ \Ket{1101} + \Ket{1110})
\nonumber
\\
\Ket{\phi_4} = &\frac{1}{\sqrt{2}}(\Ket{0110} - \Ket{1011})
\end{align}
This is the closest analytical state to the one obtained numerically. The values of the numerical
result are given in Appendix B.

The noise tolerance is defined as the maximal $p$, such that
\begin{equation}
\rho(p) = (1-p) \rho + p\frac{\mathbb{1}}{2^N},
\end{equation}
has still all the desired properties. For the  numerical state in Appendix B it is
given by $p_{max} \approx 0.35\%$.

\subsection{Five qubits}
\label{sec:level2}
We show here the five-qubit state labeled as 5a. This state is pure and 
uniquely determined by its known nearest-neighbor marginals as well. It 
is given by:
\begin{equation}\label{eq12}
\begin{split}
\Ket{\psi_5} = & \frac{1}{\mathcal{N}_1}(\frac{1}{4}\Ket{\phi}+\frac{3}{2}\sqrt{\frac{3}{10}}\Ket{\eta}+\frac{1}{14}e^{\frac{2\pi i}{3}}\Ket{11101} \\
& + \frac{2}{11}e^{\frac{-3 \pi i}{13}} \Ket{11111}),
\end{split}
\end{equation}
where $\mathcal{N}_1$ is a normalization and the corresponding substates are
\begin{equation}
\begin{split}
&\Ket{\eta} = \sqrt{\frac{2}{15}}(\frac{\sqrt{6}}{2}\Ket{\eta_1} + \sqrt{2}\Ket{\eta_2} - 2\Ket{\eta_3})\\
&\hspace{1cm} \Ket{\eta_1} = \sqrt{\frac{1}{6}}(\Ket{00011} - \Ket{00001} - \Ket{01001} +\Ket{01010} \\
&\hspace{1cm}  - \Ket{10000} + \Ket{10011}) \\
&\hspace{1cm} \Ket{\eta_2} = \frac{1}{3 \sqrt{2}}(e^{\frac{\pi i}{8}} \Ket{01100} + e^{\frac{\pi i}{8}} \Ket{11001} + 4 \Ket{11110}) \\
&\hspace{1cm} \Ket{\eta_3} = \frac{1}{2}(\Ket{11010} + \Ket{11100} + \Ket{11000} + \Ket{00000})
\end{split}
\end{equation}
and
\begin{equation}
\begin{split}
&\Ket{\phi} = \frac{1}{\mathcal{N}_2}(\frac{1}{2}\Ket{\phi_1} + \frac{\sqrt{2}}{3}e^{\frac{\pi i}{4}}\Ket{\phi_2} + \frac{\sqrt{3}}{5}\Ket{\phi_3} + 2 \Ket{\phi_4})\\
&\hspace{1cm} \Ket{\phi_1} = \frac{1}{2}(e^{\frac{\pi i}{4}}\Ket{00010} - \frac{1}{2} \Ket{01011}- \frac{1}{2}\Ket{01101} \\
&\hspace{1.7cm}+\Ket{01110} - \frac{1}{2}\Ket{01111} -\frac{1}{2}\Ket{10111} + \Ket{11011}) \\
&\hspace{1cm} \ \Ket{\phi_2} = \frac{1}{\sqrt{2}}(\Ket{00110} + \Ket{01000}) \\
&\hspace{1cm} \ \Ket{\phi_3} = \frac{1}{\sqrt{3}}(\Ket{00100} + \Ket{00101} + \Ket{10101}) \\
&\hspace{1cm} \ \Ket{\phi_4} = \frac{1}{2}(\Ket{10010} + \Ket{10100} + \Ket{10110} - \Ket{10001}) \\
\end{split}
\end{equation}
The numerically obtained state is given in Appendix B.

Due to the severe limitation on the information regarding the marginals, 
the numerically obtained state has a very low noise robustness 
($p_{max} \approx 0.11\%$), so only for little noise the entanglement 
of the global state can be proven from the separable known marginals only.

\subsection{Six Qubits}
\label{sec:level2} 
In this case, there are six possible nearest-neighbor configurations, 
five of which are pure and uniquely determined.  
Due to its size, we present one example of the obtained states the state 
in numeric form in Appendix B. The noise tolerance for this numerical
six-qubit state is $p_{max} \approx 0.02\%$, and the expectation value 
of the optimal witness, together with this state, while negative, is of 
the order of $10^{-4}$. All other marginal configurations that have been 
tested are presented in Table I.

\begin{figure}[t]
  \centering
    \includegraphics[width=0.45\textwidth]{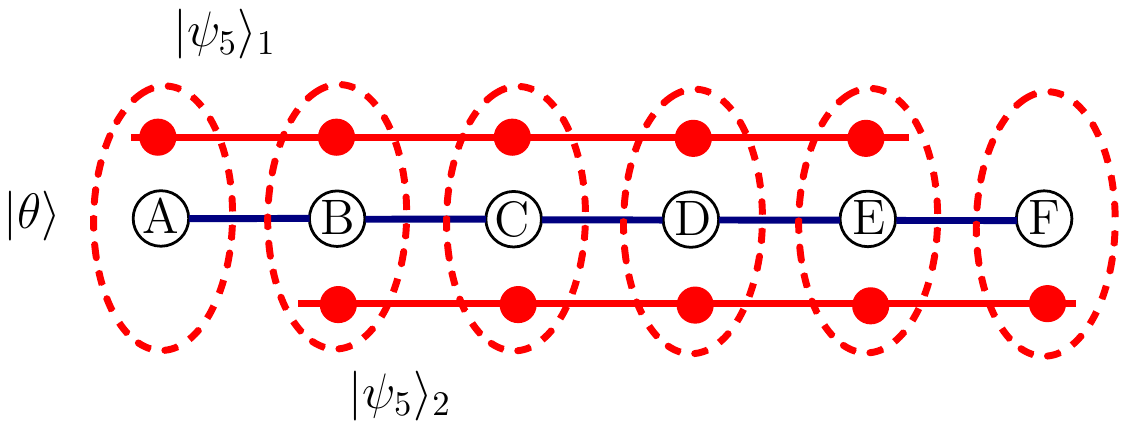}
    \caption{Constructing a state with the desired properties for a simple six-party configuration. Here, it
    is assumed that the marginals $\rho_{AB}$, $\rho_{BC}$, $\rho_{CD}$, $\rho_{DE}$, and $\rho_{EF}$ are known.
    See text for further details. }
  \label{fig:simpleex}
\end{figure}

\section{Generalization to more particles}
\label{sec:level123} 

The purpose of this last section is to present a general method for 
constructing arbitrarily large states by using the numerically found 
ones as building blocks. The main idea of this generalization was
already presented in Ref.~\cite{nikolai}. This method ensures that 
the constructed states retain the properties of the states found for 
a small number of particles. Note that we found pure genuinely multiparticle 
entangled states which are uniquely determined by their known two-body 
marginals. It is important to stress that without this unique determination 
and the purity the proposed method would not work.

Before describing the method in detail, we point out a disadvantage of
this generalization, namely that one must resort to higher-dimensional 
systems, where each party does not consist of only a qubit. The state 
we use here to exemplify is the linear five-qubit state. The first example 
is for a simple six-party system, depicted in Fig.~\ref{fig:simpleex}:

The state we want to construct is $\Ket{\theta}$. The parties A and F 
have a single qubit while the ones from B to E have two qubits. We depict 
the two-party marginals by blue lines. All we need to do is to distribute 
two copies of the pure five-qubit state $\Ket{\psi_5}$ in Eq.~{\ref{eq12}})
as represented by the thick red lines  among the six parties. 
Let us explain why this construction works: Every two-party marginal is a 
direct product of separable states since the marginals of $\Ket{\psi_5}$
are separable. Thus it is itself separable. {From} fact that $\Ket{\psi_5}$ 
is uniquely determined by its known two-body nearest-neighbor marginals it follows
that one also knows the state $\Ket{\psi_5}$ and the way copies of this state have 
been distributed among the parties. This also means that the global state itself, 
namely $\Ket{\theta}$, is uniquely determined by its two-body nearest-neighbor 
marginals as well. The constructed state is pure and cannot be factorized for 
any bipartition of the system, so it is genuinely multiparticle entangled and 
this entanglement is proven from the nearest-neighbor two-body marginals only.

\begin{figure}[t]
  \centering
    \includegraphics[width=0.35\textwidth]{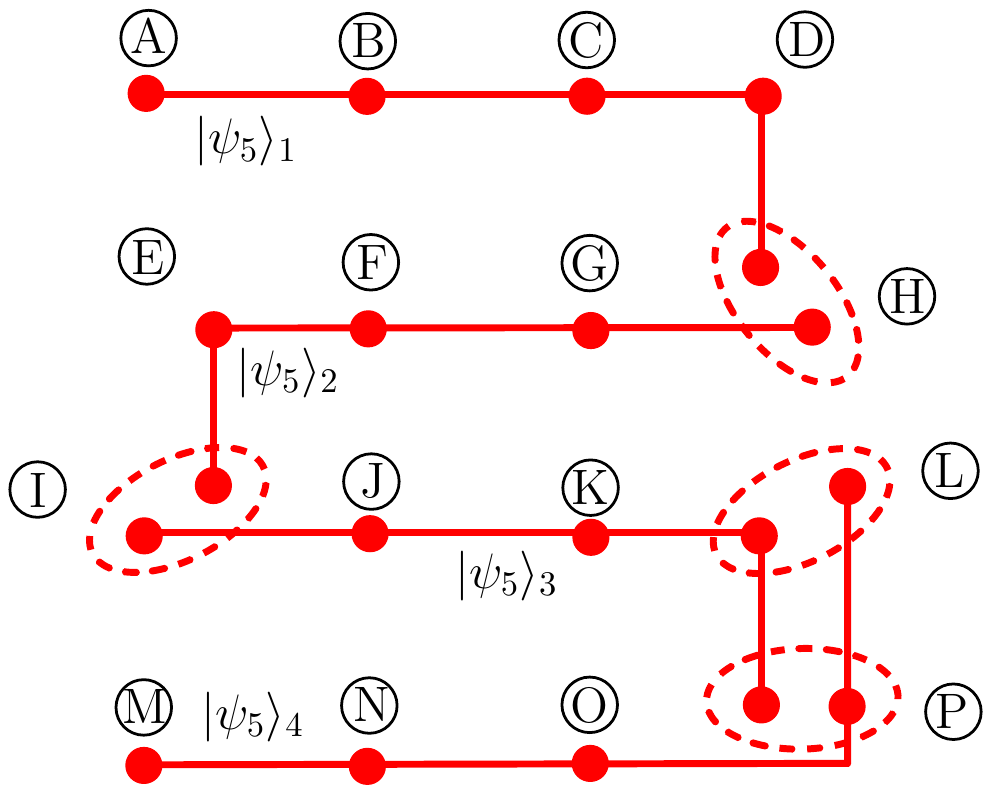}
    \caption{Covering a two-dimensional lattice with linear numerically-found states. See text for
    further details.}
  \label{fig:lattice}
\end{figure}

If we consider a $4 \times 4$ two-dimensional lattice, as depicted in Fig.~\ref{fig:lattice}, it 
can be fully covered by using four copies of the linear five-qubit state. Nodes H, I, L and 
P contain two qubits, one from each copy of the state, so we require a total of 20 qubits 
to construct the desired lattice state. 

\begin{figure}[t]
  \centering
    \includegraphics[width=0.30\textwidth]{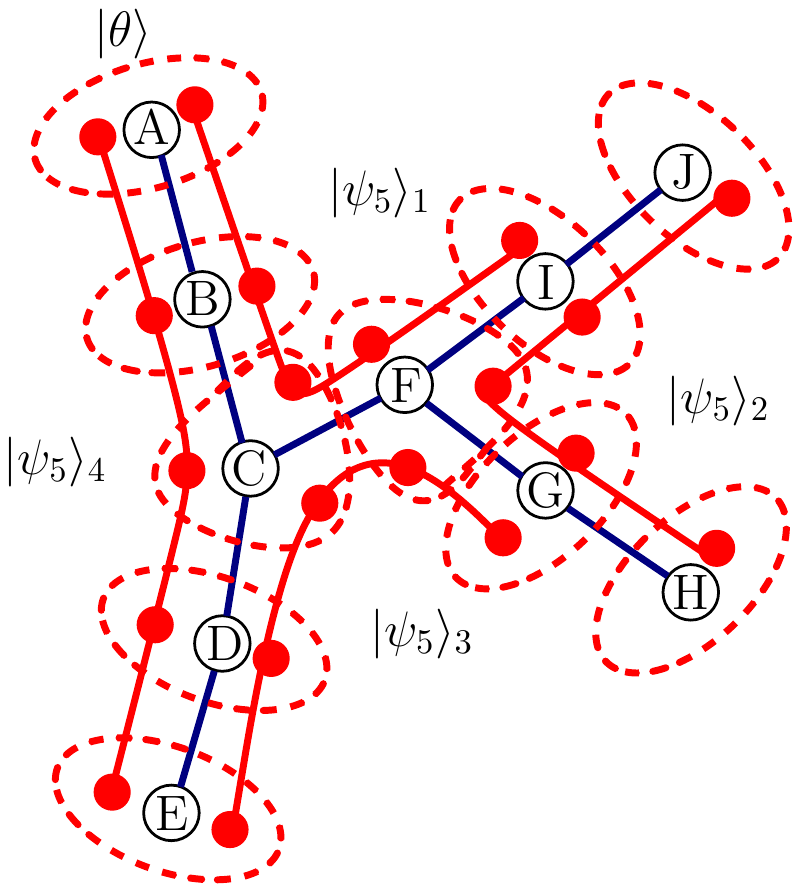}
    \caption{Constructing a state with the desired properties for an arbitrary configuration. See the text for
    further details.}
  \label{fig:uglyex}
\end{figure}

These ideas can be used for an arbitrary configuration. Consider the ten parties (A to J) in 
Fig.~\ref{fig:uglyex}. Each of the parties has at least one qubit (H and J) and at most three 
(C and F). By repeating the same algorithm as above, and distributing as few copies of the 
five-qubit state as possible, while still covering every party by at least one copy, we 
obtain a ten-particle state, with the desired properties as the one above. 
Due to the fact that with our SDP we were able to go as high as six qubits, 
one could take advantage of this and use the linear six-qubit state instead. While less robust 
to noise, an arbitrary configuration would require less copies of the state, thus helping 
to reduce the dimensionality of some of the systems (for example C and F in 
Fig.~\ref{fig:uglyex} are three-qubit systems).

\section{Conclusion}
\label{sec:level1} 

In this paper we considered an interesting class of states, namely 
genuinely multiparticle entangled states whose two-body reduced density 
matrices are all separable, nevertheless one can prove the global entanglement
from some marginals only.  We found examples of this phenomenon for all 
possible configurations of four, five and six qubits. We also showed
how these examples can be used to find more general states for more particles. 

While this paper only looks at two-body marginals, the problem could be taken one step 
further by also treating higher-order marginals and proving the entanglement of the global 
state just with the knowledge of those marginals. This gives more information about 
the state, but on the other hand, the condition on the separability of the marginals 
becomes more restrictive. Another option would be to further constrain the witness by 
allowing the observers to perform only a subset of possible measurements (for example 
only $\sigma_z$ and $\sigma_x$ measurements). Similarly, it would be interesting to
consider non-locality and ask whether all the phenomena observed for entanglement
can also be found for non-local correlations. Here, the recent results in 
Refs.~\cite{baccari, fadel} may be useful.

As previously discussed, the states presented here have a rather low noise 
tolerance (for example 0.02\% for the six qubit state), and it decreases with the 
increase in the number of qubits. This is mainly due to the strong constraint 
on nearest-neighbor information only. Nevertheless it would be interesting, 
though challenging, to see an experimental study of the effects discussed here.

\section*{Acknowledgements}
We thank Nikolai Wyderka for help with the numerical calculations.
This work has been supported by the DAAD, the ERC (Consolidator Grant 
683107/TempoQ), and the DFG.

\section*{Appendix A: Semidefinite Programming}

An SDP can be formulated as the problem of minimizing a variable
$x \in \mathbb{R}^m$ in the form 
\begin{equation}
\label{145}
\begin{split}
    \min \hspace{0.2cm}& c^T x \\
    \rm{subject\;\;to \hspace{0.1cm} } & F(x) = F_0 + \sum_{i = 1}^{m} x_i F_i \\
    & F(x) \geq 0. \\
\end{split}
\end{equation}
The vector $c \in \mathbb{R}^m$ and the $m+1$ symmetric matrices 
$F_0, ..., F_m \in \mathbb{R}^{n \times n}$ represent the problem data, 
while the $F(x) \geq 0$ constraint means that $F(x)$ is a positive 
semidefinite matrix,
$
z^T F(x) z \geq 0,
$
or, alternately, all eigenvalues of $F(x)$ are non-negative. 

An SDP is a convex optimization problem since for $F(x) \geq 0$ 
and $F(y) \geq 0$, for all $\lambda \in [0,1]$ we have
\begin{equation}
\label{147}
F[\lambda x + (1 - \lambda y)] = \lambda F(x) + (1- \lambda)F(y) \geq 0,
\end{equation}
hence both the objective function and the constraint are convex.

To a given SDP one can associate the so-called dual semidefinite program 
(from now on it will be referred to as SDD), which is of the form
\begin{equation}
\label{149}
\begin{split}
    \max \hspace{0.2cm}&  -\Tr(F_0 Z) \\
    \rm{subject\;\;to \hspace{0.1cm} } & \Tr(F_i Z) = c_i, \\
    & Z \geq 0, \\
\end{split}
\end{equation}
again, for all $i = 1,..., m$. In this case, the variable is the matrix 
$Z = Z^T \in \mathbb{R}^{n \times n}.$
Henceforth we refer to the original SDP as the {primal problem} 
and to the SDD as the {dual problem} and call a matrix $Z$ 
to be {dual feasible} if $\Tr(F_i Z) = c_i$ and $Z \geq 0.$

One important property of SDPs and their associated duals is that 
one sets bounds on the optimal value of the other. If $Z$ is dual 
feasible and $x$ is primal feasible, then we have
\begin{align}
\label{153}
c^T x & + \Tr(Z F_0) = \sum_{i = 1}^{m} \Tr(Z F_i x_i) + \Tr(Z F_0) 
\nonumber
\\
& = \Tr(Z F(x)) \geq 0,
\end{align}
since $\Tr(AB) \geq 0$ if $A \geq 0$ and $B \geq 0$. This reduces to
\begin{equation}
\label{154}
- \Tr(F_0 Z) \leq c^T x,
\end{equation}
so the dual objective value of any dual feasible point $Z$ is smaller than 
or equal to the primal objective value of any primal feasible point $x$.
If $\alpha$ is the optimal value of the SDP
$
\alpha = \min \{ c^T x | F(x) \geq 0 \},
$
then we have for any dual feasible $Z$  that 
$-\Tr(Z F_0) \leq \alpha$. Analogously if $\beta$ is the optimal value 
of the SDD, then $\beta \leq c^T x$. This means that dual feasible matrices 
impose a lower bound on the primal problem and primal feasible points impose 
an upper bound on the dual problem. What one can generally prove is that, in most cases, 
the strong condition $\alpha  = \beta$ holds:

{\bf Theorem 2.}
{\it 
The condition $\alpha = \beta$ holds if any of the following requirements is true:
\begin{itemize}
\item The primal problem is strictly feasible, that is, there exists feasible $x$ such that $F(x) > 0$.
\item The dual problem is strictly feasible, that is there exists feasible $Z$ such that $Z = Z^T > 0$.
\end{itemize}
}
A proof of this theorem can be found in Ref.~\cite{nesterov}.

\onecolumngrid
\section*{Appendix B: Examples of States}

We present here the numerical form for some of the states discussed in Section IV. The coefficients
of the states have been approximated by fractions and the states are written in a not normalized way. 

The four qubit state from Section IV for the configuration 4b is given by:
\begin{equation*}
\mathrm{\Ket{\psi_4}} \sim \biggl(
\begin{minipage}[t]{.85\displaywidth}
\mathchardef\comma=\mathcode`,
\begingroup\lccode`~=`,\lowercase{\endgroup\def~}{\comma\penalty 0 }
\mathcode`,=\string"8000
\thinmuskip=12mu plus 6mu minus 3mu \medmuskip=4mu \binoppenalty=10000
\linespread{2}\selectfont
$\displaystyle
-\frac{2}{33}+\frac{5 i}{38},\frac{1}{21}+\frac{2
   i}{13},\frac{50}{149}+\frac{3 i}{35},-\frac{1}{17}+\frac{3
   i}{25},\frac{7}{26}-\frac{5 i}{28},-\frac{3}{32}+\frac{10
   i}{61},\frac{7}{62}+\frac{6
   i}{17},-\frac{2}{17}+\frac{i}{111},\frac{5}{23}+\frac{8
   i}{29},\frac{3}{31}+\frac{5 i}{41},\frac{2}{41}-\frac{5
   i}{34},-\frac{1}{13}+\frac{11
   i}{30},-\frac{1}{289}+\frac{i}{51},-\frac{1}{270}+\frac{i}{24},-\frac{
   1}{58}+\frac{7 i}{24},\frac{11}{31}\biggr).$
\end{minipage}.
\end{equation*}

The numerical form of the five-qubit state for configuration 5a, presented in Section IV
is given by:
\begin{equation*}
\mathrm{\Ket{\psi_5}} \sim \biggl(
\begin{minipage}[t]{.85\displaywidth}
\mathchardef\comma=\mathcode`,
\begingroup\lccode`~=`,\lowercase{\endgroup\def~}{\comma\penalty 0 }
\mathcode`,=\string"8000
\thinmuskip=12mu plus 6mu minus 3mu \medmuskip=4mu \binoppenalty=10000
\linespread{2}\selectfont
$\displaystyle
-\frac{3}{35}-\frac{i}{22},\frac{4}{31}+\frac{i}{40},\frac{4}{29}-
   \frac{i}{22},-\frac{1}{28}+\frac{4 i}{29},\frac{4}{35}-\frac{2
   i}{25},-\frac{1}{24}+\frac{i}{19},-\frac{6}{35}-\frac{5
   i}{28},\frac{2}{33}-\frac{4
   i}{45},\frac{1}{32}-\frac{i}{3},\frac{3}{35}-\frac{19
   i}{94},-\frac{5}{24}-\frac{3 i}{16},\frac{1}{74}-\frac{2
   i}{33},-\frac{4}{27}+\frac{i}{207},-\frac{1}{186}-\frac{2
   i}{39},\frac{5}{41}-\frac{2 i}{13},\frac{2}{19}+\frac{5
   i}{34},-\frac{2}{27}+\frac{i}{6},-\frac{3}{29}-\frac{8
   i}{33},-\frac{1}{8}-\frac{5 i}{36},\frac{7}{30}-\frac{3
   i}{40},-\frac{4}{31}-\frac{5 i}{28},\frac{1}{7}-\frac{3
   i}{35},-\frac{11}{36}+\frac{i}{83},\frac{1}{50}-\frac{2
   i}{35},\frac{1}{10}-\frac{8
   i}{41},\frac{1}{26}-\frac{i}{50},-\frac{4}{39}-\frac{2
   i}{29},-\frac{2}{29}-\frac{2
   i}{19},-\frac{1}{18}-\frac{i}{295},-\frac{2}{27}-\frac{2
   i}{23},-\frac{1}{18}-\frac{4 i}{33},\frac{1}{10}\biggr).$
\end{minipage}
\end{equation*}

Due to its size, we only present the linear six-qubit state (configuration 6a) in numerical form:
\begin{equation*}
\mathrm{\Ket{\psi_6}} \sim \biggl(
\begin{minipage}[t]{.85\displaywidth}
\mathchardef\comma=\mathcode`,
\begingroup\lccode`~=`,\lowercase{\endgroup\def~}{\comma\penalty 0 }
\mathcode`,=\string"8000
\thinmuskip=12mu plus 6mu minus 3mu \medmuskip=4mu \binoppenalty=10000
\linespread{2.0}\selectfont
$\displaystyle
\frac{3}{77},0,\frac{1}{55},-\frac{4}{61}-\frac{6 i}{85},-\frac{2}{97},0,0,\frac{3}{107},-\frac{10}{59},\frac{1}{133}-\frac{7i}{167},-\frac{7}{211},-\frac{7}{211},-\frac{1}{27}+\frac{21 i}{106},-\frac{23}{172}-\frac{6 i}{121},-\frac{11}{70},\frac{1}{106}-\frac{8i}{193},-\frac{7}{211},\frac{9}{167},\frac{13}{168},-\frac{33}{97},\frac{10}{103}+\frac{17 i}{77},\frac{13}{168},\frac{8}{61},\frac{8}{83},\frac{16}{43},0,\frac{1}{128}-\frac{11i}{84},\frac{29}{83},0,\frac{47}{168},\frac{16}{43},\frac{9}{167},0,0,-\frac{1}{117},-\frac{17}{358}+\frac{15 i}{188},-\frac{1}{79}+\frac{10i}{191},0,0,-\frac{7}{181}+\frac{i}{25},\frac{2}{63}+\frac{9 i}{103},0,0,\frac{10}{91},\frac{5}{127}+\frac{9 i}{79},0,\frac{2}{59}+\frac{5i}{72},-\frac{3}{137},-\frac{3}{107},-\frac{6}{161}+\frac{4 i}{59},-\frac{8}{159}+\frac{4 i}{93},\frac{28}{159}+\frac{29i}{217},-\frac{3}{137},-\frac{3}{107},-\frac{3}{107},-\frac{5}{62}+\frac{2 i}{17},\frac{3}{137},0,\frac{3}{137},-\frac{3}{107},-\frac{2}{43}-\frac{7i}{87},\frac{1}{174},0,-\frac{3}{65}+
\frac{6 i}{125}\biggr).$
\end{minipage}
\end{equation*}
This state is also pure and uniquely determined by its two-body nearest-neighbor marginals, while still retaining the desired properties, namely the state should be genuinely multiparticle entangled and the entanglement should be proven from the separable nearest-neighbor two-body marginals only. 


\twocolumngrid

\end{document}